\documentclass[12pt]{article} 
\usepackage{amsmath, amsfonts, amsthm, hyperref, fullpage}

\title{Quantum interpolation of polynomials}
\author{Daniel M. Kane\thanks{Dept.\ of Mathematics, Harvard University, One Oxford Street, Cambridge MA 02138.  Email:\ {\tt dankane@math.harvard.edu}} \and Samuel A. Kutin\thanks{IDA/CCR-P, 805 Bunn Drive, Princeton NJ 08540.  Email:\ {\tt kutin@idaccr.org}}}
\date{August 2009}

\newtheorem{theorem}{Theorem}

\newcommand{\qubit}[1]{\left|{#1}\right\rangle}
\newcommand{\FF}{{\cal F}}
\newcommand{\PP}{{\cal P}}
\newcommand{\eps}{\epsilon}
\newcommand{\E}{\mathop{\mathbf E}\limits}
\newcommand{\qhat}{\hat{q}}

\theoremstyle{remark}
\newtheorem*{remark}{Remark}

\begin{document}
\maketitle
\begin{abstract}
We consider quantum interpolation of polynomials.  We
imagine a quantum computer with
black-box access to input/output pairs $(x_i, f(x_i))$,
where $f$ is a degree-$d$ polynomial, and we wish to
compute $f(0)$.  We give asymptotically tight quantum lower
bounds for this problem, even in the case where $0$ is
among the possible values of $x_i$.
\end{abstract}

\section{Introduction}

Can a quantum computer efficiently interpolate polynomials?
Can it distinguish low-degree from high-degree polynomials?
We consider black-box algorithms that seek to learn information
about a
polynomial $f$ from input/output pairs $(x_i, f(x_i))$.
We define a more general class of
\emph{$(d,S)$-independent} function properties, where, outside
of a set $S$ of exceptions, knowing $d$ input values does not help one
predict the answer.  There are essentially two strategies
to computing such a function:  query $d+1$ random input
values, or search for one of the $|S|$ exceptions.  We
show that, up to constant factors, we cannot beat these
two approaches.

Let $\FF$ be a collection of functions from some domain
$D$ to some range $R$.  A \emph{property} is a (nontrivial) map
$\PP \colon \FF \to \{0, 1\}$.  We say that $\PP$
is \emph{$(d,S)$-independent} for some subset $S \subseteq D$
if, for any $z_1$, $\ldots$, $z_d \in D$ with
$z_1$, $\ldots$, $z_r \in S$ and $z_{r+1}$, $\ldots$, $z_d \notin S$
(where $r = |\{z_i\} \cap S|$),
the $(d-r)$-tuple of values $(f(z_{r+1}), \ldots, f(z_d))$ is
independent of the $(r+1)$-tuple $(f(z_1), \ldots, f(z_r), \PP(f))$.
We say that $\PP$ is \emph{$d$-independent} if it is
$(d,\emptyset)$-independent.
(For simplicity, we consider independence with respect to
the uniform distribution on $\FF$.)

For example, let $\FF$ be the set of degree-$d$ polynomials
from some finite field $K$ to itself.  Let $R = K$,
and let $D$ be some subset of $K$.  We could define $\PP(f)$
to be one bit of information about a particular function
value $f(z)$.  If $z \notin D$, then this property is
$d$-independent; knowing any $d$ values of a degree-$d$
polynomial yields no information about any other value.
If $z \in D$, then $\PP$ is $(d,\{z\})$-independent.
Alternatively, we could define $\PP(f)$ to be one bit of information
about a (nonconstant) coefficient of $f$; this is also
$d$-independent.

We analyze quantum algorithms that compute such a $\PP$ based
on black-box access to the function $f$.  We consider
two models:

\begin{itemize}
\item In the \emph{chosen-input} model, we give our oracle
$x \in D$ and it returns $f(x)$.  (More precisely, the
oracle transformation maps $\qubit{x, b, c}$ to
$\qubit{x, b + f(x), c}$, where $+$ is some appropriate
reversible notion of addition.)
\item In the \emph{random-input} model, there is some
map $X$ from $\{1, .., n\}$ onto $D$.
We give our oracle $i$ and it
returns the pair $(X(i), Y(i))$, where $Y(i) = f(X(i))$.
This oracle is our only access to the map $X$.
\end{itemize}

The random-input model may seem unusual.  It is a natural
extension of Valiant's PAC learning model~\cite{Val84}
to the quantum setting, although it differs slightly from the
quantum PAC model introduced by Bshouty and
Jackson~\cite{BJ}.  For technical reasons, we consider
distributions on maps $X$ with the same image $D$.  We
say that such a distribution is \emph{permutation-independent}
if, for any permutation $\sigma$ on $D$, the maps $X$ and
$\sigma \circ X$ have the same probability.

We prove a result for each model.  We say that the
\emph{bias} of an algorithm is its edge over random
guessing; that is, on any input $f$, the algorithm outputs
$\PP(f)$ with probability at least $\frac12 + \eps$.

\begin{theorem}\label{thm-main-chosen}
Let $\PP$ be a $d$-independent property of
a family of functions $\FF$.  Let
$A$ be a quantum query algorithm, in the chosen-input
model, which, for any $f \in \FF$, correctly
computes $\PP(f)$ with positive bias.  Then
the number of queries made by $A$ is at least $(d+1)/2$.
\end{theorem}

\begin{theorem}\label{thm-main-random}
Let $\PP$ be a $(d,S)$-independent property of
a family of functions $\FF$ with a domain of size $n$.  Let
$\Delta$ be a permutation-independent distribution of
random maps.  Let
$A$ be a quantum query algorithm, in the random-input
model, which, for any $f \in \FF$, and with $X \sim \Delta$,
computes $\PP(f)$ with bias at least $\eps$.  Then
the number of queries made by $A$ is at least
$$
\min\left\{\frac{d+1}{2}, C_\eps \sqrt{\frac{n}{|S|}}\right\},
$$
where $C_\eps$ is a constant depending on $\eps$.
\end{theorem}

To return to our first example, suppose $\FF$ is
the set of degree-$d$ polynomials, and $\PP(f)$ is one
bit of information about $f(z)$ for some $z \notin D$.
One strategy is to make $d+1$ queries to compute $d+1$
different values of $f(X(i))$, interpolate the polynomial,
and read off $f(z)$.  The above theorems show that, for
either query model, this approach is within a factor
of $2$ of being optimal.

What if, instead, $z \in D$?  In the chosen-input
model, computing $f(z)$ is no longer interesting; we
can perform a single query.  In the random-input model,
we could still query $d+1$ points and interpolate, or
we could use Grover search~\cite{Grover} to find the
value of $i$ with $X(i) = z$, at which point one additional
query gives the answer.  \autoref{thm-main-random}
says that one of these two strategies must be optimal,
up to a constant factor.

We survey lower bound methods in \autoref{sec-prior},
focusing on the approach we will use:\ the
polynomial method~\cite{BBCMdW}.  We then prove
the two above theorems in \autoref{sec-proofs} and
give some final thoughts in \autoref{sec-conclude}.

\section{Lower Bound Methods}\label{sec-prior}

There are several standard techniques for proving
quantum query lower bounds.  One approach is to use
information theory.  For example, suppose our goal were
not to compute $f(z)$ at a single point, but to
produce a complete description of the degree-$d$
polynomial $f$.  This requires specifying $d+1$
coefficients, each an element of $K$.  But each query
gives us, information-theoretically, at most two elements of $K$.
By an interactive version of
Holevo's Theorem~\cite[Theorem 2]{CvDNT}, we require
at least $(d+1)/4$ queries.  However, this approach
does not apply to computing a single value $f(z)$.

A second approach is to use the ``adversary'' method
of Ambainis~\cite{Ambainis}.  The basic idea, in our
setting, would be to find a collection of functions
$g \in \FF$ with $\PP(g) = 0$, and another collection
of $h \in \FF$ with $\PP(h) = 1$, where each $g$ is
``close to'' many $h$, in the sense that they agree
on almost all inputs.  However, any two distinct
polynomials disagree on almost all inputs.  H{\o}yer,
Lee, and \v{S}palek~\cite{HLS}, after noting that
Ambainis's original method cannot prove a non-constant lower
bound when $0$-inputs and $1$-inputs disagree on a
constant fraction of the inputs, propose a variant with
``negative weights'' that, in theory, does not run up
against this barrier.  In practice, even this generalized adversary
method has not yet yielded a nonconstant lower bound for such
a problem.

We will apply the polynomial method~\cite{BBCMdW}.
For the chosen-input model, let $\delta_{x,y}$ be
the function of $f$ that is $1$ when $f(x) = y$ and
$0$ otherwise.  Then the quantum query maps
$$
\qubit{x, b, c} \mapsto \sum_{y} \delta_{x,y} \qubit{x, b + y, c}.
$$
So, if we start in some fixed state, after a single
query each amplitude is an affine expression in the
values $\delta_{x,y}$.  After $T$ queries, each amplitude
is a polynomial in $\{\delta_{x,y}\}$ of degree at most $T$.
We now measure the state and output some bit; the probability
that this bit is $1$ is thus a polynomial of
degree at most $2T$.  This polynomial $p$ satisfies the following
properties:
\begin{itemize}
\item If $\delta_{x,y}$ encodes any function from $D$ to $R$
(that is, each $\delta_{x,y} \in \{0,1\}$
and $\sum_{y\in R} \delta_{x,y} = 1$ for all $x$),
then $0 \le p(\{\delta_{x,y}\}) \le 1$.
\item If $\delta_{x,y}$ encodes some $f \in \FF$, then
$|p(\{\delta_{x,y}\}) - \PP(f)| < \frac12$.
\end{itemize}

A lower bound on the degree of such
a polynomial thus
gives a lower bound on the number of quantum queries.

For the random-input model, the same idea applies;
the variable $\delta_{i,x,y}$ is $1$ when $X(i) = x$ and
$f(x) = y$ and $0$ otherwise, and
$$
\qubit{i, a, b, c} \mapsto \sum_{x,y} \delta_{i,x,y} \qubit{i, a+x, b+y, c}.
$$
The polynomial $p$ in this setting satisfies the properties:
\begin{itemize}
\item If $\delta_{i,x,y}$ encodes any functions $X$ from indices
to $D$ and $Y$ from indices to $R$
(that is, each $\delta_{i,x,y} \in \{0,1\}$
and $\sum_{x,y} \delta_{i,x,y} = 1$ for all $i$),
then $0 \le p(\{\delta_{i,x,y}\}) \le 1$.
\item If $\delta_{i,x,y}$ encodes $X$ and $f \circ X$ for
some $f \in \FF$, then
$|p(\{\delta_{i,x,y}\}) - \PP(f)| \le \frac12 - \eps$.
\end{itemize}

In early uses of the polynomial method~\cite{BBCMdW},
one step in a typical application was to symmetrize down
to a polynomial in one variable.  This works well for
total functions, but not for promise problems.  (Here, the
promise is that $f$ represent some function.)
The method has been adapted to a similar setting for proving
a lower bound for the element distinctness problem~\cite{AS,K};
in this case, symmetrizing separately on the domain and
range yields a function of two variables.  We will use
a similar approach to tackle interpolation.

\begin{remark}
There are different ways to prove lower bounds on the degree of
a polynomial computing a function.  For example, a referee for
an early version of this paper noted that \autoref{thm-main-chosen}
above can be proved using a general result\footnote{See the discussion
following their Lemma 3~\cite{BVdW}.} of Buhrman, et al.~\cite{BVdW}.
We give a direct proof whose main idea
generalizes to the random-input problem.
\end{remark}

\section{Proofs}\label{sec-proofs}

We now prove our main results.  We begin with
the chosen-input model.

\begin{proof}[Proof of \autoref{thm-main-chosen}]
Let $A$ be an algorithm
computing the $d$-independent property $\PP$
with nonzero bias.  Suppose that
$A$ makes fewer than $(d+1)/2$ queries.  As discussed
in \autoref{sec-prior}, we write the probability that
$A$ outputs $1$ as a polynomial $p(f)$, by which we mean
a polynomial in the variables $\{\delta_{x,y}\}$, of
degree at most $d$.
When $f \in \PP^{-1}(0)$, then
$0 \le p(f) < \frac12$; when
$f \in \PP^{-1}(1)$, then
$\frac12 < p(f) \le 1$.

Write $p$ as a sum of monomials $\sum_k m_k$.
Each monomial has the form
$$m_k = \prod_{j=1}^t \delta_{x_j,y_j}$$
for some $t \le d$.  Hence, each $m_k$ depends
on at most $d$ values of $f$.  By the definition
of $d$-independence, the expected value of $m_k$
over $\PP^{-1}(0)$ is the same as it is over $\PP^{-1}(1)$.
This is true for all $k$, so
$$
\frac12 < \E_{f \in \PP^{-1}(1)} [p(f)]
= \E_{f \in \PP^{-1}(0)} [p(f)] < \frac12.
$$
This is impossible.  We conclude that no such algorithm
exists; that is, any algorithm computing $\PP$ requires
at least $(d+1)/2$ queries.
\end{proof}

The proof of \autoref{thm-main-random} is more involved.
We will first show that, assuming an algorithm makes fewer
than $(d+1)/2$ queries, the actual values of $f(x)$ do
not matter unless $x$ is in the special set $S$.
This first part of the argument uses the same logic
as the proof of \autoref{thm-main-chosen}.

Intuitively, if the values $f(x)$ matter only for $x \in S$,
the simplest possible case would be one where any such value
of $f(x)$ immediately yields the answer $\PP(f)$.  This is
Grover search, with a known lower bound of $\Omega(\sqrt{n/|S|})$.
The second part of the proof of \autoref{thm-main-random}
represents one approach to formalizing this intuition.

\begin{proof}[Proof of \autoref{thm-main-random}]
Let $A$ be an algorithm
computing the $(d,S)$-independent property $\PP$
with bias at least $\eps$.  Suppose that
$A$ makes fewer than $(d+1)/2$ queries.  As discussed
in \autoref{sec-prior}, we write the probability that
$A$ outputs $1$ as a polynomial $p(X,Y)$, by which we mean
a polynomial in the variables $\{\delta_{i,x,y}\}$, of
degree at most $d$.  For any $i$ and any $x \notin S$, we introduce
the variables $\xi_{i,x}$ (which is $1$ when $X(i) = x$
and $0$ otherwise) and $\upsilon_{i,y}$ (which is $1$
when $Y(i) = y$ and $0$ otherwise), and we write
$\delta_{i,x,y} = \xi_{i,x} \upsilon_{i,y}$.
For all $X \colon \{1, \ldots, n\} \to D$ and $Y \colon \{1, \ldots, n\} \to R$,
we have $0 \le p(X, Y) \le 1$; we will use this generality.
When $f \in \FF$, we have
$|p(X,f \circ X) - \PP(f)| \le \frac12 - \eps$.

Write $p$ as a sum of monomials $\sum_k m_k$.
Each monomial has the form
$$
m_k = \prod_{j=1}^r \delta_{i_j, x_j, y_j}
\prod_{j=r+1}^t \xi_{i_j, x_j} \upsilon_{i_j, y_j}
$$
for some $r \le t \le d$ with $x_j \in S$ for $j \le r$
and $x_j \notin S$ for $j > r$, and with all $i_j$
distinct.  By the definition of
$d$-independence, once we condition on $X(i_j) = x_j$ for $1 \le j \le t$,
the expected value of $\prod_{j=r+1}^t \upsilon_{i_j,y_j}$
over $f \in \FF$ and $X \sim \Delta$
is independent of $\PP(f)$ and of the values $\delta_{i_j, x_j, y_j}$
for $j \le r$.  Hence, we can replace this product with its
expected value over $f$ and $X$, yielding a new polynomial $q$ using only the
variables $\delta_{i,x,y}$ (for $x \in S$) and $\xi_{i,x}$
(for $x \notin S$).  The polynomial $q$ satisfies the
original conditions:  $0 \le q(X,Y) \le 1$ for any $X, Y$,
and $|q(X, f \circ X) - \PP(f)| \le \frac12 - \eps$ when $f \in \FF$.
Furthermore, $\deg q \le \deg p$.

If $S = \emptyset$, then $q$ depends only on $X$ but not $f$,
which is impossible.  In this case, $A$ must have made at least
$(d+1)/2$ queries.  For the remainder of the proof we assume $S$
is nonempty.

We now apply $q$ to a particular set of instances.
Let $k = |S|$, write $S = \{z_1, \ldots, z_k\}$,
and write $D \setminus S = \{z_{k+1}, \ldots, z_n\}$.
We will permute these values in blocks.  Let
$B = \left\lfloor n/k \right\rfloor$.  For any
function $\pi$ from $\{0, \ldots, B-1\}$ to $\{0, \ldots, B-1\}$
we get an arrangement given by $X(i + kj) =
z_{i + k\pi(j)}$ for $1 \le i \le k$ and $0 \le j < B$.
(We write $X(i) = z_i$ for $i > Bk$.)  When $\pi$ is a
permutation, the list $\{X(i)\}$ covers all of $D$.

Now, choose some $g, h \in \FF$ with $\PP(g) = 0$ and
$\PP(h) = 1$.  We let $Y(i + kj)$ (where $1 \le i \le k$)
be $g(z_i)$ when $j$ is even and $h(z_i)$ when $j$ is odd.
Fixing these values, any function $\pi$ gives us values
of $\delta_{i,x,y}$ (for $x \in S$) and $\xi_{i,x}$ (for
$x \notin S$).  We let $q(\pi)$ denote the result of applying
the polynomial $q$ to these values.  It is clear that we can
rewrite each $\xi_{i,x}$ or $\delta_{i,x,y}$ as $0$ or
as some $\eta_{i,j}$, which is defined to be
$1$ if $\pi(i) = j$ and $0$ otherwise.  Hence, $q(\pi)$
is a polynomial in $\{\eta_{i,j}\}$.

For any function $\pi$, we must have $0 \le q(\pi) \le 1$.
For a permutation $\pi$ with $\pi^{-1}(0)$ even, we have
$q(\pi) = q(X, g \circ X) \le \frac12 - \eps$.  For a permutation $\pi$
with $\pi^{-1}(0)$ odd, we have $q(\pi) = q(X, h \circ X) \ge \frac12 + \eps$.
We have reduced to the standard problem of permutation inversion; as
first shown by Ambainis~\cite{Ambainis}, we
know that any such polynomial has degree $\Omega(\sqrt{B})$.

For concreteness, we finish the proof using symmetrization.
First, we symmetrize $q$ with respect to any rearrangement
of the values $1, \ldots, B-1$ in the range of $\pi$.  This
reduces us to variables $\{\eta_i\}$ where $\eta_i = 1$ when
$\pi(i) = 0$ and $0$ otherwise.  Next, we symmetrize with
respect to any rearrangement of even $i$ and any rearrangement
of odd $i$.  We are left with a polynomial $q(\alpha, \beta)$
in two variables:  $\alpha$
counts the number of even $i$ with $\pi(i) = 0$, and
$\beta$ counts the number of odd $i$ with $\pi(i) = 0$.

Note that $0 \le q(\alpha, \beta) \le 1$ for any
$0 \le \alpha \le \lceil B/2 \rceil$ and any
$0 \le \beta \le \lfloor B/2 \rfloor$.  Furthermore,
$q(1,0) \le \frac12 - \eps$ and $q(0,1) \ge \frac12 + \eps$.
We break into two cases depending on whether $q(0,0)$ is
at least $\frac12$ or at most $\frac12$.  In either case,
we get a polynomial $\qhat$ in one variable with $0 \le \qhat(i) \le 1$
for $i = 0, \ldots, \lfloor B/2\rfloor$ and with a constant
gap between $\qhat(0)$ and $\qhat(1)$.  By a lemma
of Paturi~\cite{Pat92} (see also \cite{BBCMdW, NS}),
we conclude that $\deg \qhat = \Omega(\sqrt{B})$
as desired.  By construction, $\deg \qhat \le \deg q \le \deg p$.
\end{proof}

\section{Conclusions}\label{sec-conclude}

We have proven a lower bound of $(d+1)/2$ for polynomial
interpolation (to find $f(z)$ when $z$ is not in
the domain of queries).  The usual classical algorithm,
of course, requires $d+1$ queries.  We suspect that
a quantum algorithm should also require $d+1$ queries,
but we do not have a proof.

It is worth noting that, in the generality in which
it is stated, \autoref{thm-main-chosen} is
tight.  Let $\FF$ be the set of all functions from
some domain to $\{0,1\}$, and let $\PP(f)$ be 
the parity $\bigoplus_{x \in U} f(x)$ of some
collection of input places with $|U| = d+1$.  This
is a $d$-independent property; any set of
$d$ values, even if they all lie in $U$, are
independent of the final answer.  In this case,
the standard Deutsch--Josza algorithm~\cite{DJ}
computes the parity with $(d+1)/2$ queries.
\autoref{thm-main-chosen} can be viewed as
an extension of the parity lower bound of
Farhi, et al.~\cite{FGGS}.
Hence, any stronger lower bound for polynomial
interpolation would require using some additional
structure of the problem.

The authors' original proof of \autoref{thm-main-chosen}
did not use the polynomial method.  Instead,
following the same general lines as Ambainis's
proof of the adversary lower bound~\cite{Ambainis},
we kept track of density matrices.  If, after some
number of queries, we cannot distinguish $0$-inputs
from $1$-inputs even given $m$ additional classical
queries, then after one more quantum query we cannot
distinguish $0$-inputs from $1$-inputs given $m-2$
additional queries.  The initial value of $m$ is $d$,
so if we make fewer than $(d+1)/2$ queries the
final value is at least $0$, meaning that we cannot
gain any information about the answer.

The authors moved away from this proof, both because
it was harder to formalize and because it did not
adapt as well to \autoref{thm-main-random}.  However,
it may be that combining this original idea with the
adversary method could lead to even stronger bounds
on similar problems.

\section*{Acknowledgments}
The authors thank Bruce Jordan for
proposing this problem.  We also thank Bob Beals
and David Moulton for helpful discussions. 
\bibliography{inter}
\bibliographystyle{alpha}

\end{document}